# Photoelectric Observations, Light Curves Analysis and Period Study of the Eclipsing Variable DO Cas

Shahin Jafarzadeh

*Biruni Observatory, Shiraz University, Shiraz 71454, Iran*

**Abstract**

The new B and V photoelectric observations of the β Lyrae eclipsing binary DO Cas were obtained on 6 nights form December 2000 to January 2001. The observations were made at the Biruni Observatory, Shiraz, Iran and the light curves are analyzed using the Wilson light curve synthesis and differential correction code. So, the relative surface luminosities, new light elements, and new orbital elements have been obtained, and from times of minimum the period is improved. With these and previously published times, the period variation is studied and a constant period is approved, though some authors has mentioned some variations. The solutions of the light curves suggest that DO Cas is a contact binary.

## 1.  Introduction

The variability of DO Cas was discovered in 1940 by Hoffmeister (1947). In his 1947 paper, Hoffmeister concluded the β Lyrae type of the variable from a series of visual observations. The star's J2000 coordinates are   R.A.= $02^h41^m24^s$,   Dec.= +60°33'11" , and it has the following catalogue designations: HD 16506 , BD +59°529 , SAO 12388 , HIC 12543 , Pickering 023360.

The spectral type of the primary was classified to be A2 (Henry Draper Catalogue), A2II (Fehrenbach *et al*., 1966), and A4-A5 (Hill *et al*., 1975). Also the spectral type of the secondary was estimated as F2 by Koch (1973).

Concerning the nature of the system, Schneller and Daene (1952), Cester *et al*. (1977), Kaluzny (1985), Rovithis-Livaniou and Rovithis (1986), Oh and Ahn (1992), Kim *et al*. (1999), and Oh and Oh (2000) have concluded that it is a contact binary while Giannone and Giannuzzi (1974), and Liu *et al*. (1988) have specified a semi-detached classification.

The results of the present analysis support the contact binary classification.

This eclipsing binary has been observed by many observers such as Winkler (1966), Kukarkin *et al*. (1969), Gleim and Winkler (1969), Aluigi *et al*. (1995), Oh and Kim (1996), and the others mentioned above. A summery of the spectroscopic elements for DO Cas is given by Kranjc (1960) and the single-line spectroscopic data were obtained by Mannino (1958), who mentioned that the orbit of the system has an eccentricity of 0.12. However, Kaluzny (1985) commented that such a large eccentricity is suspicious.

In this study, I am presenting the light curves of DO Cas in the two Johnson filters B and V. I analyzed these light curves using the Wilson program, and the parameters derived from observed light curves are compared with the previous studies.

The purpose of the present work is to present new photoelectric times of minimum light, to make a period study of this system and to derive new light elements.

## 2. Observations

A total of 473 blue and 474 yellow photoelectric observations of DO Cas were carried out during 6 nights from December 2000 to January 2001 with the 51 cm cassegrian telescope of Biruni Observatory, Shiraz, Iran, as listed with all the corrections in Table 1 and Table 2 along with the JD and corresponding phase value. The telescope is equipped with an un-refrigerated RCA4509 multiplier phototube operating at 1600 Volts and with the B and V filters matched to Johnson UBV system. The readings were amplified and digitized through an A/D converter and a personal computer was used to handle and reduce the data.

The observing sequence was the usual pattern of *sky – comparison – variable (4 times) – comparison – sky*, with each observation lasting about 10 seconds.

The comparison star used for the observations was BD+59°521 (HD 16088, SAO 23474, HIC 1228) which is the same one as used by Aluigi *et al*. (1995). The J2000 coordinates for this star are: R.A. = $02^h37^m26^s$, Dec. = +60°05'18", respectively. The check star wasn't observed, because no appreciable variation was detected.

All observations were corrected for differential atmospheric extinction effects using the extinction coefficients were determined from the observations of the comparison star by using a reduction computer program which is known as Redwip code, written by G.McCook.

The average values for the extinction coefficients were 0.298 in blue light and 0.196 in yellow light.

The phases have been computed using Koch *et al*. (1963) ephemeris formulae. Figure 1 represents the B and V light curves of DO Cas, as they were derived from my observations.

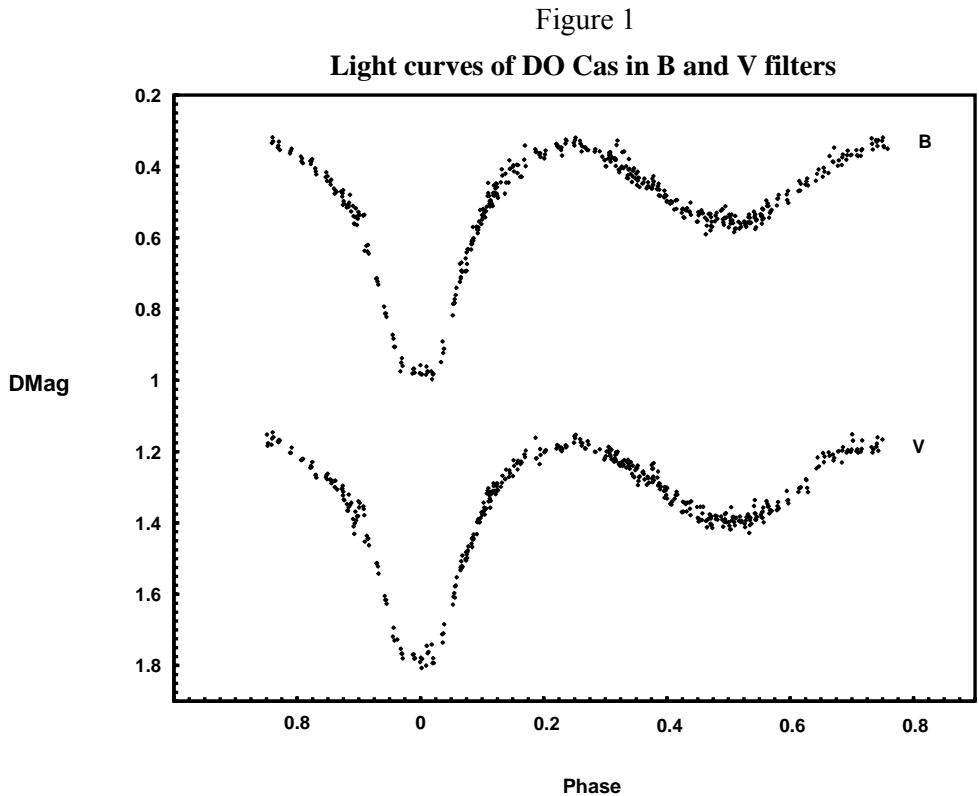

Figure 1

**Light curves of DO Cas in B and V filters**

TABLE 1
Observational points of DO Cas for the B-filter

| JD HEL 2451900+ | PHASE | DMAG ($m_v - m_c$) | JD HEL 2451900+ | PHASE | DMAG ($m_v - m_c$) | JD HEL 2451900+ | PHASE | DMAG ($m_v - m_c$) |
|---|---|---|---|---|---|---|---|---|
| 9.24645 | 0.05546 | 0.7721 | 10.22306 | 0.48187 | 0.5467 | 11.23164 | 0.95497 | 0.8724 |
| .24753 | 0.05704 | 0.7609 | .22952 | 0.49130 | 0.5567 | .23252 | 0.95625 | 0.8836 |
| .24845 | 0.05839 | 0.7409 | .23043 | 0.49263 | 0.5637 | .23348 | 0.95765 | 0.9059 |
| .25245 | 0.06422 | 0.7241 | .23125 | 0.49383 | 0.5457 | .23421 | 0.95872 | 0.9063 |
| .25342 | 0.06564 | 0.7145 | .23223 | 0.49525 | 0.5115 | .24043 | 0.96780 | 0.9749 |
| .25420 | 0.06679 | 0.6700 | .23296 | 0.49632 | 0.5431 | .24129 | 0.96906 | 0.9506 |
| .25498 | 0.06792 | 0.6957 | .23807 | 0.50379 | 0.5523 | .24202 | 0.97013 | 0.9382 |
| .25881 | 0.07352 | 0.6578 | .23878 | 0.50482 | 0.5351 | .24280 | 0.97126 | 0.9599 |
| .25995 | 0.07517 | 0.6409 | .23975 | 0.50624 | 0.5509 | .25361 | 0.98705 | 0.9807 |
| .26080 | 0.07642 | 0.6325 | .24062 | 0.50751 | 0.5586 | .25457 | 0.98845 | 0.9682 |
| .26175 | 0.07781 | 0.6325 | .24659 | 0.51623 | 0.5555 | .25544 | 0.98972 | 0.9676 |
| .26519 | 0.08283 | 0.6178 | .24766 | 0.51779 | 0.5744 | .25618 | 0.99080 | 0.9789 |
| .26603 | 0.08407 | 0.6178 | .24870 | 0.51931 | 0.5596 | .26196 | 0.99926 | 0.9802 |
| .26684 | 0.08525 | 0.5699 | .24945 | 0.52041 | 0.5694 | .26260 | 0.00019 | 0.9574 |
| .26743 | 0.08611 | 0.6016 | .25301 | 0.52562 | 0.5542 | .26352 | 0.00152 | 0.9825 |
| .27099 | 0.09130 | 0.5699 | .25373 | 0.52666 | 0.5593 | .26418 | 0.00248 | 0.9835 |
| .27186 | 0.09257 | 0.5778 | .25436 | 0.52758 | 0.5649 | .26870 | 0.00909 | 0.9844 |
| .27264 | 0.09372 | 0.5778 | .25524 | 0.52886 | 0.5695 | .26930 | 0.00997 | 0.9615 |
| .27345 | 0.09490 | 0.5544 | .25822 | 0.53322 | 0.5349 | .27033 | 0.01148 | 0.9747 |
| .27705 | 0.10016 | 0.5368 | .25889 | 0.53420 | 0.5590 | .27114 | 0.01266 | 0.9758 |
| .27823 | 0.10188 | 0.5190 | .25962 | 0.53527 | 0.5585 | .27457 | 0.01767 | 0.9754 |
| .27903 | 0.10305 | 0.5323 | .26048 | 0.53652 | 0.5578 | .27560 | 0.01917 | 0.9970 |
| .27983 | 0.10422 | 0.5228 | .26462 | 0.54257 | 0.5402 | .27635 | 0.02027 | 0.9835 |
| .28395 | 0.11023 | 0.4750 | .26527 | 0.54352 | 0.5758 | .27702 | 0.02125 | 0.9821 |
| .28491 | 0.11164 | 0.4827 | .26608 | 0.54470 | 0.5707 | .28539 | 0.03347 | 0.9489 |
| .28567 | 0.11275 | 0.4976 | .26678 | 0.54572 | 0.5489 | .28712 | 0.03599 | 0.8910 |
| .28647 | 0.11392 | 0.4980 | .27119 | 0.55216 | 0.5339 | .28789 | 0.03712 | 0.9233 |
| .28947 | 0.11830 | 0.4740 | .27184 | 0.55312 | 0.5396 | .28859 | 0.03814 | 0.9114 |
| .29022 | 0.11940 | 0.4530 | .27257 | 0.55418 | 0.5370 | .29840 | 0.05247 | 0.8176 |
| .29077 | 0.12019 | 0.4818 | .27325 | 0.55517 | 0.5188 | .29905 | 0.05342 | 0.7856 |
| .29133 | 0.12102 | 0.4748 | .27757 | 0.56149 | 0.4955 | .29975 | 0.05445 | 0.7841 |
| 10.15591 | 0.38379 | 0.4632 | .27829 | 0.56254 | 0.5111 | .30053 | 0.05558 | 0.7825 |
| .15682 | 0.38512 | 0.4641 | .27908 | 0.56369 | 0.5033 | .30666 | 0.06454 | 0.6733 |
| .15786 | 0.38664 | 0.4811 | .27977 | 0.56470 | 0.5112 | .30743 | 0.06566 | 0.7107 |
| .15941 | 0.38891 | 0.4828 | 11.16955 | 0.86427 | 0.4711 | .30810 | 0.06664 | 0.6927 |
| .16418 | 0.39587 | 0.4866 | .17444 | 0.87142 | 0.4703 | .30885 | 0.06774 | 0.6933 |
| .16504 | 0.39712 | 0.4801 | .17593 | 0.87360 | 0.4751 | .31240 | 0.07293 | 0.6944 |
| .16606 | 0.39861 | 0.4818 | .17669 | 0.87470 | 0.4665 | .31301 | 0.07381 | 0.6936 |
| .16694 | 0.39990 | 0.4994 | .18253 | 0.88324 | 0.5091 | .31365 | 0.07475 | 0.6927 |
| .17518 | 0.41193 | 0.5207 | .18343 | 0.88456 | 0.5020 | .31430 | 0.07570 | 0.6736 |
| .17660 | 0.41401 | 0.4954 | .18427 | 0.88577 | 0.4799 | .31886 | 0.08236 | 0.6332 |
| .17747 | 0.41528 | 0.5191 | .18539 | 0.88741 | 0.5261 | .31962 | 0.08346 | 0.6104 |
| .17827 | 0.41645 | 0.4942 | .18995 | 0.89408 | 0.5418 | .32031 | 0.08447 | 0.6055 |
| .18645 | 0.42840 | 0.5443 | .19077 | 0.89528 | 0.5178 | .32118 | 0.08574 | 0.6080 |
| .18746 | 0.42987 | 0.5214 | .19202 | 0.89710 | 0.5640 | .32482 | 0.09107 | 0.5690 |
| .18858 | 0.43151 | 0.5237 | .19269 | 0.89808 | 0.5396 | .32618 | 0.09304 | 0.5871 |
| .18936 | 0.43264 | 0.5171 | .20265 | 0.91262 | 0.6226 | .32696 | 0.09419 | 0.5706 |
| .20387 | 0.45384 | 0.5311 | .20355 | 0.91394 | 0.6384 | .32877 | 0.09683 | 0.5222 |
| .20542 | 0.45610 | 0.5228 | .20431 | 0.91505 | 0.6202 | .33258 | 0.10239 | 0.5377 |
| .20620 | 0.45724 | 0.5311 | .20514 | 0.91625 | 0.6446 | .33386 | 0.10427 | 0.5277 |
| .20693 | 0.45830 | 0.5310 | .21341 | 0.92834 | 0.7147 | .33446 | 0.10515 | 0.5432 |
| .21336 | 0.46770 | 0.5335 | .21412 | 0.92937 | 0.7142 | .33532 | 0.10640 | 0.5256 |
| .21424 | 0.46898 | 0.5335 | .21482 | 0.93040 | 0.7229 | .33934 | 0.11226 | 0.5056 |
| .21524 | 0.47044 | 0.5335 | .21567 | 0.93164 | 0.7317 | .33988 | 0.11306 | 0.5122 |
| .21599 | 0.47154 | 0.5417 | .22227 | 0.94127 | 0.7933 | .34045 | 0.11389 | 0.4947 |
| .22053 | 0.47816 | 0.5539 | .22327 | 0.94274 | 0.8127 | .34098 | 0.11466 | 0.5013 |
| .22115 | 0.47908 | 0.5376 | .22402 | 0.94384 | 0.8122 | .34589 | 0.12183 | 0.4617 |
| .22219 | 0.48060 | 0.5299 | .22477 | 0.94492 | 0.8219 | .34656 | 0.12281 | 0.4478 |

TABLE 1
(Continued)

| JD HEL 2451900+ | PHASE | DMAG ($m_v-m_c$) | JD HEL 2451900+ | PHASE | DMAG ($m_v-m_c$) | JD HEL 2451900+ | PHASE | DMAG ($m_v-m_c$) |
|---|---|---|---|---|---|---|---|---|
| 11.34714 | 0.12366 | 0.4491 | 12.26932 | 0.47057 | 0.5797 | 15.14541 | 0.67128 | 0.3475 |
| .34771 | 0.12449 | 0.4581 | .27005 | 0.47163 | 0.5672 | .14964 | 0.67747 | 0.3982 |
| 12.15583 | 0.30480 | 0.3781 | .27067 | 0.47253 | 0.5553 | .15059 | 0.67886 | 0.3834 |
| .15700 | 0.30651 | 0.3617 | .27130 | 0.47346 | 0.5694 | .15397 | 0.68379 | 0.3960 |
| .15807 | 0.30808 | 0.3597 | .28101 | 0.48764 | 0.5245 | .15455 | 0.68464 | 0.3743 |
| .15917 | 0.30969 | 0.3726 | .28260 | 0.48996 | 0.5202 | .15527 | 0.68568 | 0.3670 |
| .16356 | 0.31609 | 0.3780 | .28317 | 0.49079 | 0.5483 | .15919 | 0.69142 | 0.3813 |
| .16437 | 0.31728 | 0.3808 | .28510 | 0.49361 | 0.5447 | .16006 | 0.69268 | 0.3688 |
| .16515 | 0.31841 | 0.3400 | .28962 | 0.50022 | 0.5712 | .16064 | 0.69353 | 0.3560 |
| .16615 | 0.31988 | 0.3271 | .29026 | 0.50115 | 0.5353 | .16378 | 0.69811 | 0.3690 |
| .17078 | 0.32664 | 0.4011 | .29107 | 0.50233 | 0.5434 | .16481 | 0.69961 | 0.3673 |
| .17148 | 0.32766 | 0.3606 | .29180 | 0.50340 | 0.5604 | .17020 | 0.70749 | 0.3547 |
| .17250 | 0.32915 | 0.4047 | .29565 | 0.50903 | 0.5845 | .17087 | 0.70847 | 0.3693 |
| .17337 | 0.33043 | 0.3583 | .29641 | 0.51013 | 0.5750 | .17212 | 0.71030 | 0.3705 |
| .17782 | 0.33692 | 0.4335 | .29701 | 0.51101 | 0.5390 | .17501 | 0.71452 | 0.3707 |
| .17864 | 0.33812 | 0.3796 | .29773 | 0.51205 | 0.5742 | .17564 | 0.71544 | 0.3551 |
| .17954 | 0.33944 | 0.4076 | .31079 | 0.53114 | 0.5574 | .17667 | 0.71694 | 0.3525 |
| .18042 | 0.34072 | 0.4212 | .31165 | 0.53239 | 0.5433 | .18706 | 0.73212 | 0.3216 |
| .18574 | 0.34848 | 0.4338 | .31238 | 0.53346 | 0.5570 | .18762 | 0.73293 | 0.3367 |
| .18684 | 0.35009 | 0.4203 | .31306 | 0.53445 | 0.5439 | .18849 | 0.73420 | 0.3526 |
| .18767 | 0.35131 | 0.4527 | .31775 | 0.54130 | 0.5272 | .19206 | 0.73942 | 0.3262 |
| .18831 | 0.35224 | 0.4152 | .31843 | 0.54230 | 0.5446 | .19280 | 0.74051 | 0.3312 |
| .19844 | 0.36705 | 0.4615 | .31918 | 0.54340 | 0.5348 | .19353 | 0.74157 | 0.3432 |
| .19930 | 0.36830 | 0.4569 | .32025 | 0.54495 | 0.5520 | .19440 | 0.74284 | 0.3269 |
| .20009 | 0.36945 | 0.4448 | .32653 | 0.55413 | 0.5607 | .19889 | 0.74940 | 0.3287 |
| .20092 | 0.37066 | 0.4562 | .32716 | 0.55504 | 0.5498 | .19978 | 0.75070 | 0.3192 |
| .20422 | 0.37548 | 0.4598 | .32781 | 0.55599 | 0.5208 | .20043 | 0.75165 | 0.3432 |
| .20487 | 0.37643 | 0.4354 | .32844 | 0.55692 | 0.5462 | .20107 | 0.75258 | 0.3463 |
| .20556 | 0.37744 | 0.4509 | .33442 | 0.56564 | 0.5443 | .20479 | 0.75802 | 0.3504 |
| .20621 | 0.37839 | 0.4425 | .33548 | 0.56720 | 0.5304 | .20543 | 0.75895 | 0.3343 |
| .21096 | 0.38532 | 0.4560 | .33623 | 0.56830 | 0.5270 | .20607 | 0.75988 | 0.3185 |
| .21160 | 0.38627 | 0.4485 | .33732 | 0.56989 | 0.5222 | .20679 | 0.76093 | 0.3299 |
| .21261 | 0.38774 | 0.4656 | .34310 | 0.57832 | 0.5101 | .21183 | 0.76830 | 0.3473 |
| .21326 | 0.38868 | 0.4826 | .34369 | 0.57918 | 0.4911 | .21249 | 0.76926 | 0.3422 |
| .21744 | 0.39479 | 0.4858 | .34444 | 0.58028 | 0.4901 | .21328 | 0.77041 | 0.3305 |
| .21813 | 0.39580 | 0.4684 | .34506 | 0.58119 | 0.4802 | .21403 | 0.77151 | 0.3537 |
| .22005 | 0.39861 | 0.4819 | .35413 | 0.59443 | 0.4793 | .22547 | 0.78821 | 0.3635 |
| .22105 | 0.40006 | 0.4970 | .35451 | 0.59499 | 0.4948 | .22611 | 0.78916 | 0.3569 |
| .22458 | 0.40522 | 0.5051 | .35553 | 0.59648 | 0.4688 | .22711 | 0.79061 | 0.3507 |
| .22519 | 0.40611 | 0.4982 | .35590 | 0.59702 | 0.5028 | .23763 | 0.80598 | 0.3722 |
| .22585 | 0.40708 | 0.4998 | .36708 | 0.61335 | 0.4694 | .23829 | 0.80694 | 0.3858 |
| .22652 | 0.40806 | 0.5014 | .36790 | 0.61455 | 0.4410 | .23889 | 0.80782 | 0.3780 |
| .23745 | 0.42401 | 0.5307 | .36881 | 0.61588 | 0.4682 | .23976 | 0.80909 | 0.3915 |
| .23800 | 0.42483 | 0.5210 | .36965 | 0.61710 | 0.4490 | .24808 | 0.82124 | 0.3853 |
| .23866 | 0.42579 | 0.5193 | .37518 | 0.62518 | 0.4509 | .24867 | 0.82211 | 0.3932 |
| .23928 | 0.42669 | 0.5433 | .37585 | 0.62616 | 0.4424 | .24931 | 0.82304 | 0.3865 |
| .24572 | 0.43610 | 0.5274 | .37673 | 0.62745 | 0.4342 | .24999 | 0.82403 | 0.3798 |
| .24640 | 0.43708 | 0.5274 | .37760 | 0.62871 | 0.4534 | .25351 | 0.82917 | 0.4059 |
| .24707 | 0.43806 | 0.5021 | 15.12522 | 0.64180 | 0.4404 | .25415 | 0.83010 | 0.4143 |
| .24768 | 0.43896 | 0.5362 | .12653 | 0.64371 | 0.4210 | .25468 | 0.83088 | 0.4225 |
| .25410 | 0.44834 | 0.5554 | .12812 | 0.64603 | 0.4086 | .26519 | 0.84623 | 0.4301 |
| .25501 | 0.44966 | 0.5550 | .13278 | 0.65284 | 0.4293 | .26592 | 0.84729 | 0.4166 |
| .25568 | 0.45064 | 0.5288 | .13387 | 0.65443 | 0.4044 | .26651 | 0.84816 | 0.4413 |
| .25622 | 0.45143 | 0.5371 | .13477 | 0.65575 | 0.4154 | .26711 | 0.84903 | 0.4274 |
| .26317 | 0.46158 | 0.5441 | .13922 | 0.66224 | 0.4191 | .27003 | 0.85329 | 0.4426 |
| .26407 | 0.46290 | 0.5907 | .14021 | 0.66369 | 0.3711 | .27055 | 0.85406 | 0.4515 |
| .26515 | 0.46447 | 0.5669 | .14405 | 0.66930 | 0.3864 | .27116 | 0.85495 | 0.4373 |
| .26575 | 0.46535 | 0.5511 | .14470 | 0.67025 | 0.3879 | .27426 | 0.85948 | 0.4621 |

TABLE 1
(Continued)

| JD HEL 2451900+ | PHASE | DMAG ($m_v-m_c$) | JD HEL 2451900+ | PHASE | DMAG ($m_v-m_c$) | JD HEL 2451900+ | PHASE | DMAG ($m_v-m_c$) |
|---|---|---|---|---|---|---|---|---|
| 15.27507 | 0.86067 | 0.4771 | 16.15454 | 0.14519 | 0.4072 | 16.23843 | 0.26772 | 0.3470 |
| .27589 | 0.86187 | 0.4680 | .15781 | 0.14996 | 0.4018 | .24033 | 0.27049 | 0.3541 |
| .28356 | 0.87306 | 0.4849 | .15844 | 0.15089 | 0.3966 | .24153 | 0.27225 | 0.3539 |
| .28420 | 0.87400 | 0.5084 | .15913 | 0.15189 | 0.4145 | .24881 | 0.28288 | 0.3588 |
| .28503 | 0.87520 | 0.4983 | .16222 | 0.15640 | 0.3955 | .24949 | 0.28386 | 0.3519 |
| .28572 | 0.87622 | 0.5052 | .16287 | 0.15735 | 0.4187 | .25423 | 0.29079 | 0.3576 |
| .28859 | 0.88041 | 0.5155 | .16350 | 0.15828 | 0.3962 | .25729 | 0.29525 | 0.3728 |
| .28926 | 0.88139 | 0.5117 | .16616 | 0.16216 | 0.4290 | .25996 | 0.29916 | 0.3828 |
| .28989 | 0.88230 | 0.5082 | .16684 | 0.16314 | 0.4294 | .26094 | 0.30060 | 0.3993 |
| .29045 | 0.88313 | 0.5136 | .16755 | 0.16419 | 0.4298 | .26170 | 0.30169 | 0.3999 |
| .29549 | 0.89048 | 0.5110 | .17061 | 0.16865 | 0.3809 | .26498 | 0.30650 | 0.3683 |
| .29604 | 0.89130 | 0.5396 | .17125 | 0.16958 | 0.3416 | .26564 | 0.30746 | 0.3664 |
| .29669 | 0.89224 | 0.5602 | .17199 | 0.17067 | 0.3992 | .26653 | 0.30876 | 0.3869 |
| .29728 | 0.89311 | 0.5625 | .18276 | 0.18640 | 0.3523 | .26719 | 0.30972 | 0.3772 |
| .30092 | 0.89841 | 0.5467 | .18346 | 0.18742 | 0.3597 | .27037 | 0.31436 | 0.3779 |
| .30152 | 0.89929 | 0.5282 | .18410 | 0.18837 | 0.3595 | .27092 | 0.31517 | 0.3795 |
| .30214 | 0.90021 | 0.5368 | .18734 | 0.19310 | 0.3673 | .27149 | 0.31600 | 0.3811 |
| .30275 | 0.90108 | 0.5367 | .18790 | 0.19391 | 0.3627 | .27222 | 0.31706 | 0.3987 |
| .30679 | 0.90698 | 0.5383 | .18854 | 0.19484 | 0.3738 | .27545 | 0.32178 | 0.4094 |
| .30729 | 0.90773 | 0.5372 | .19273 | 0.20096 | 0.3778 | .27609 | 0.32272 | 0.4167 |
| .30788 | 0.90859 | 0.5360 | .19343 | 0.20199 | 0.3667 | .27677 | 0.32370 | 0.4081 |
| .30845 | 0.90942 | 0.6375 | .19416 | 0.20306 | 0.3556 | .28388 | 0.33410 | 0.3997 |
| 16.12269 | 0.09867 | 0.5621 | .20620 | 0.22064 | 0.3371 | .28453 | 0.33505 | 0.4324 |
| .12351 | 0.09987 | 0.5496 | .20690 | 0.22167 | 0.3425 | .28550 | 0.33647 | 0.4250 |
| .12450 | 0.10131 | 0.5439 | .20752 | 0.22256 | 0.3408 | .28634 | 0.33769 | 0.4093 |
| .12876 | 0.10753 | 0.5133 | .20807 | 0.22338 | 0.3467 | .29013 | 0.34323 | 0.4438 |
| .12966 | 0.10885 | 0.4833 | .21083 | 0.22740 | 0.3474 | .29086 | 0.34429 | 0.4283 |
| .13040 | 0.10993 | 0.4465 | .21141 | 0.22824 | 0.3547 | .29163 | 0.34541 | 0.4131 |
| .13446 | 0.11586 | 0.5058 | .21196 | 0.22906 | 0.3470 | .29240 | 0.34654 | 0.4061 |
| .13538 | 0.11720 | 0.5113 | .21250 | 0.22983 | 0.3697 | .29647 | 0.35248 | 0.4402 |
| .13609 | 0.11825 | 0.4812 | .21623 | 0.23529 | 0.3360 | .29707 | 0.35336 | 0.4322 |
| .13957 | 0.12332 | 0.4764 | .21673 | 0.23602 | 0.3249 | .29783 | 0.35447 | 0.4325 |
| .14042 | 0.12457 | 0.4822 | .21736 | 0.23693 | 0.3279 | .29847 | 0.35540 | 0.4577 |
| .14112 | 0.12558 | 0.4886 | .22547 | 0.24878 | 0.3338 | .30174 | 0.36019 | 0.4483 |
| .14444 | 0.13043 | 0.4449 | .22615 | 0.24978 | 0.3414 | .30239 | 0.36113 | 0.4468 |
| .14540 | 0.13184 | 0.4079 | .22675 | 0.25066 | 0.3264 | .30310 | 0.36216 | 0.4450 |
| .14615 | 0.13294 | 0.4262 | .22747 | 0.25171 | 0.3191 | .30390 | 0.36333 | 0.4348 |
| .14887 | 0.13691 | 0.4765 | .23238 | 0.25888 | 0.3283 | .31186 | 0.37496 | 0.4625 |
| .14946 | 0.13777 | 0.4206 | .23299 | 0.25977 | 0.3358 | .31247 | 0.37586 | 0.4622 |
| .15020 | 0.13885 | 0.4456 | .23414 | 0.26144 | 0.3584 | .31342 | 0.37724 | 0.4530 |
| .15312 | 0.14311 | 0.4453 | .23475 | 0.26234 | 0.3434 | .31401 | 0.37810 | 0.4272 |
| .15379 | 0.14409 | 0.3833 | .23788 | 0.26691 | 0.3471 | | | |

TABLE 2
Observational points of DO Cas for the V-filter

| JD HEL 2451900+ | PHASE | DMAG $(m_v-m_c)$ | JD HEL 2451900+ | PHASE | DMAG $(m_v-m_c)$ | JD HEL 2451900+ | PHASE | DMAG $(m_v-m_c)$ |
|---|---|---|---|---|---|---|---|---|
| 9.24697 | 0.05622 | 1.5782 | 10.22335 | 0.48229 | 1.3908 | 11.23287 | 0.95676 | 1.6938 |
| .24795 | 0.05766 | 1.5758 | .23078 | 0.49314 | 1.3966 | .23370 | 0.95797 | 1.7298 |
| .24890 | 0.05905 | 1.5527 | .23156 | 0.49427 | 1.3756 | .23692 | 0.96267 | 1.7270 |
| .25297 | 0.06498 | 1.5249 | .23248 | 0.49563 | 1.3726 | .24068 | 0.96817 | 1.7534 |
| .25383 | 0.06625 | 1.5079 | .23321 | 0.49669 | 1.3980 | .24151 | 0.96939 | 1.7666 |
| .25460 | 0.06736 | 1.5204 | .23836 | 0.50421 | 1.3560 | .24228 | 0.97050 | 1.7671 |
| .25539 | 0.06851 | 1.5231 | .23917 | 0.50540 | 1.3928 | .24302 | 0.97158 | 1.7805 |
| .25930 | 0.07423 | 1.4814 | .24018 | 0.50687 | 1.3833 | .25405 | 0.98769 | 1.7695 |
| .26039 | 0.07582 | 1.4984 | .24092 | 0.50795 | 1.3832 | .25493 | 0.98898 | 1.7685 |
| .26125 | 0.07708 | 1.4771 | .24697 | 0.51679 | 1.3945 | .25569 | 0.99009 | 1.7807 |
| .26196 | 0.07812 | 1.4660 | .24806 | 0.51838 | 1.3818 | .25642 | 0.99116 | 1.7799 |
| .26567 | 0.08354 | 1.4429 | .24907 | 0.51985 | 1.3974 | .26221 | 0.99961 | 1.7908 |
| .26644 | 0.08466 | 1.4423 | .25002 | 0.52124 | 1.3945 | .26289 | 0.00061 | 1.7788 |
| .26706 | 0.08557 | 1.4417 | .25331 | 0.52606 | 1.4018 | .26375 | 0.00186 | 1.8067 |
| .26764 | 0.08642 | 1.4321 | .25392 | 0.52693 | 1.4116 | .26441 | 0.00282 | 1.7814 |
| .27142 | 0.09193 | 1.3973 | .25476 | 0.52817 | 1.3932 | .26891 | 0.00940 | 1.8002 |
| .27226 | 0.09316 | 1.3976 | .25543 | 0.52915 | 1.3842 | .26955 | 0.01033 | 1.7451 |
| .27301 | 0.09426 | 1.3890 | .25842 | 0.53351 | 1.4279 | .27056 | 0.01182 | 1.7658 |
| .27388 | 0.09553 | 1.3982 | .25928 | 0.53476 | 1.4065 | .27134 | 0.01295 | 1.7620 |
| .27789 | 0.10138 | 1.3710 | .26007 | 0.53593 | 1.3856 | .27512 | 0.01848 | 1.7410 |
| .27864 | 0.10247 | 1.3427 | .26081 | 0.53701 | 1.4024 | .27587 | 0.01956 | 1.7931 |
| .27942 | 0.10362 | 1.3571 | .26485 | 0.54291 | 1.3356 | .27655 | 0.02056 | 1.7798 |
| .28026 | 0.10484 | 1.3629 | .26550 | 0.54386 | 1.3906 | .27722 | 0.02154 | 1.7930 |
| .28439 | 0.11088 | 1.3394 | .26631 | 0.54504 | 1.3914 | .28648 | 0.03506 | 1.7124 |
| .28529 | 0.11220 | 1.3224 | .26704 | 0.54610 | 1.3830 | .28735 | 0.03633 | 1.7360 |
| .28607 | 0.11333 | 1.3140 | .27139 | 0.55246 | 1.3764 | .28813 | 0.03748 | 1.7098 |
| .28670 | 0.11426 | 1.3139 | .27202 | 0.55337 | 1.3830 | .28882 | 0.03848 | 1.6843 |
| .28984 | 0.11884 | 1.2961 | .27279 | 0.55451 | 1.3707 | .29862 | 0.05279 | 1.6288 |
| .29044 | 0.11972 | 1.3118 | .27347 | 0.55549 | 1.3680 | .29926 | 0.05372 | 1.6070 |
| .29098 | 0.12050 | 1.3112 | .27781 | 0.56183 | 1.3384 | .30005 | 0.05489 | 1.5972 |
| .29155 | 0.12134 | 1.3105 | .27853 | 0.56289 | 1.3463 | .30074 | 0.05589 | 1.6097 |
| 10.15638 | 0.38448 | 1.2730 | .27930 | 0.56401 | 1.3542 | .30696 | 0.06498 | 1.5326 |
| .15727 | 0.38578 | 1.2559 | .27998 | 0.56500 | 1.3533 | .30764 | 0.06596 | 1.5295 |
| .15845 | 0.38750 | 1.2895 | 11.16985 | 0.86471 | 1.3058 | .30833 | 0.06698 | 1.5262 |
| .16004 | 0.38982 | 1.2977 | .17518 | 0.87250 | 1.3046 | .30911 | 0.06811 | 1.4914 |
| .16462 | 0.39651 | 1.2993 | .17619 | 0.87398 | 1.2960 | .31262 | 0.07325 | 1.5019 |
| .16559 | 0.39793 | 1.3090 | .17691 | 0.87502 | 1.3128 | .31321 | 0.07411 | 1.5053 |
| .16650 | 0.39925 | 1.3012 | .18290 | 0.88378 | 1.3209 | .31389 | 0.07509 | 1.4883 |
| .16734 | 0.40049 | 1.3020 | .18374 | 0.88500 | 1.3486 | .31451 | 0.07600 | 1.4815 |
| .17554 | 0.41246 | 1.3319 | .18451 | 0.88613 | 1.3502 | .31927 | 0.08295 | 1.4585 |
| .17694 | 0.41450 | 1.3220 | .18560 | 0.88772 | 1.3437 | .31992 | 0.08390 | 1.4674 |
| .17775 | 0.41568 | 1.3126 | .19025 | 0.89451 | 1.3683 | .32068 | 0.08501 | 1.4463 |
| .17857 | 0.41688 | 1.3208 | .19106 | 0.89570 | 1.3935 | .32142 | 0.08610 | 1.4451 |
| .18675 | 0.42884 | 1.3436 | .19227 | 0.89746 | 1.3820 | .32508 | 0.09144 | 1.4320 |
| .18781 | 0.43037 | 1.3465 | .19290 | 0.89839 | 1.3806 | .32657 | 0.09362 | 1.3914 |
| .18891 | 0.43198 | 1.3677 | .20291 | 0.91301 | 1.4359 | .32830 | 0.09614 | 1.3975 |
| .18970 | 0.43315 | 1.3427 | .20375 | 0.91423 | 1.4447 | .32923 | 0.09751 | 1.3863 |
| .20428 | 0.45443 | 1.3547 | .20457 | 0.91543 | 1.4439 | .33283 | 0.10276 | 1.3663 |
| .20575 | 0.45658 | 1.3917 | .20537 | 0.91659 | 1.4625 | .33405 | 0.10454 | 1.3767 |
| .20650 | 0.45768 | 1.3732 | .21360 | 0.92861 | 1.5134 | .33492 | 0.10581 | 1.3592 |
| .20723 | 0.45874 | 1.3551 | .21437 | 0.92974 | 1.5128 | .33556 | 0.10675 | 1.3599 |
| .21373 | 0.46824 | 1.3875 | .21526 | 0.93105 | 1.5223 | .33952 | 0.11253 | 1.3316 |
| .21469 | 0.46964 | 1.3762 | .21592 | 0.93201 | 1.5426 | .34005 | 0.11331 | 1.3575 |
| .21556 | 0.47091 | 1.4021 | .22280 | 0.94205 | 1.6047 | .34066 | 0.11419 | 1.3380 |
| .21634 | 0.47204 | 1.3818 | .22348 | 0.94305 | 1.6158 | .34121 | 0.11500 | 1.3190 |
| .22080 | 0.47857 | 1.3797 | .22428 | 0.94421 | 1.6159 | .34607 | 0.12210 | 1.3196 |
| .22143 | 0.47948 | 1.3894 | .22503 | 0.94531 | 1.6271 | .34681 | 0.12318 | 1.3132 |
| .22247 | 0.48100 | 1.3809 | .23190 | 0.95534 | 1.7190 | .34732 | 0.12393 | 1.2973 |

TABLE 2

(Continued)

| JD HEL 2451900+ | PHASE | DMAG ($m_v-m_c$) | JD HEL 2451900+ | PHASE | DMAG ($m_v-m_c$) | JD HEL 2451900+ | PHASE | DMAG ($m_v-m_c$) |
|---|---|---|---|---|---|---|---|---|
| 11.34789 | 0.12476 | 1.3078 | 12.27028 | 0.47197 | 1.4063 | 15.14996 | 0.67793 | 1.1999 |
| 12.15627 | 0.30544 | 1.1871 | .27087 | 0.47284 | 1.4055 | .15081 | 0.67918 | 1.1982 |
| .15728 | 0.30692 | 1.2038 | .27149 | 0.47373 | 1.4145 | .15415 | 0.68404 | 1.2307 |
| .15835 | 0.30849 | 1.2126 | .28123 | 0.48796 | 1.3821 | .15479 | 0.68499 | 1.1969 |
| .15952 | 0.31020 | 1.2214 | .28281 | 0.49026 | 1.3911 | .15549 | 0.68601 | 1.1876 |
| .16381 | 0.31647 | 1.2216 | .28357 | 0.49138 | 1.4105 | .15944 | 0.69177 | 1.2027 |
| .16460 | 0.31762 | 1.2012 | .28548 | 0.49417 | 1.3898 | .16025 | 0.69295 | 1.2001 |
| .16541 | 0.31880 | 1.2055 | .28989 | 0.50061 | 1.3981 | .16082 | 0.69380 | 1.1961 |
| .16641 | 0.32025 | 1.2006 | .29057 | 0.50161 | 1.4080 | .16400 | 0.69843 | 1.2008 |
| .17104 | 0.32702 | 1.2287 | .29134 | 0.50272 | 1.4180 | .16501 | 0.69992 | 1.1943 |
| .17175 | 0.32806 | 1.2406 | .29206 | 0.50377 | 1.4181 | .16564 | 0.70083 | 1.1516 |
| .17275 | 0.32952 | 1.2366 | .29587 | 0.50935 | 1.3983 | .16646 | 0.70203 | 1.1695 |
| .17363 | 0.33080 | 1.2236 | .29662 | 0.51043 | 1.4071 | .17037 | 0.70775 | 1.1989 |
| .17809 | 0.33731 | 1.2424 | .29724 | 0.51134 | 1.3867 | .17139 | 0.70923 | 1.1932 |
| .17888 | 0.33848 | 1.2410 | .29797 | 0.51241 | 1.3858 | .17231 | 0.71057 | 1.2033 |
| .17978 | 0.33978 | 1.2481 | .31126 | 0.53182 | 1.3762 | .17520 | 0.71479 | 1.1998 |
| .18070 | 0.34113 | 1.2295 | .31195 | 0.53283 | 1.3848 | .17624 | 0.71632 | 1.1923 |
| .18597 | 0.34882 | 1.2518 | .31266 | 0.53386 | 1.3835 | .17688 | 0.71725 | 1.1690 |
| .18708 | 0.35045 | 1.2452 | .31332 | 0.53482 | 1.3923 | .18723 | 0.73236 | 1.1875 |
| .18791 | 0.35166 | 1.2467 | .31795 | 0.54159 | 1.3526 | .18829 | 0.73391 | 1.2025 |
| .18853 | 0.35256 | 1.2392 | .31865 | 0.54262 | 1.3730 | .18908 | 0.73506 | 1.1941 |
| .19866 | 0.36737 | 1.2932 | .31947 | 0.54382 | 1.3738 | .19239 | 0.73990 | 1.1870 |
| .19953 | 0.36863 | 1.2705 | .32050 | 0.54532 | 1.3849 | .19302 | 0.74083 | 1.1778 |
| .20030 | 0.36975 | 1.2747 | .32673 | 0.55442 | 1.3845 | .19395 | 0.74218 | 1.1603 |
| .20113 | 0.37097 | 1.2873 | .32738 | 0.55536 | 1.3856 | .19464 | 0.74319 | 1.1978 |
| .20445 | 0.37582 | 1.2755 | .32803 | 0.55631 | 1.3569 | .19916 | 0.74979 | 1.1662 |
| .20509 | 0.37675 | 1.2859 | .32870 | 0.55729 | 1.3980 | .19999 | 0.75100 | 1.1519 |
| .20579 | 0.37778 | 1.2701 | .33498 | 0.56647 | 1.3735 | .20064 | 0.75195 | 1.1838 |
| .20646 | 0.37876 | 1.2805 | .33571 | 0.56754 | 1.3642 | .20126 | 0.75286 | 1.1768 |
| .21116 | 0.38562 | 1.2953 | .33645 | 0.56862 | 1.3649 | .20500 | 0.75832 | 1.1810 |
| .21192 | 0.38672 | 1.2929 | .33761 | 0.57031 | 1.3660 | .20564 | 0.75925 | 1.1637 |
| .21283 | 0.38806 | 1.2900 | .34333 | 0.57866 | 1.3639 | .20636 | 0.76030 | 1.1464 |
| .21349 | 0.38902 | 1.3057 | .34394 | 0.57955 | 1.3527 | .20723 | 0.76157 | 1.1594 |
| .21765 | 0.39509 | 1.3106 | .34468 | 0.58064 | 1.3412 | .21205 | 0.76862 | 1.1719 |
| .21843 | 0.39624 | 1.3295 | .34541 | 0.58170 | 1.3590 | .21277 | 0.76967 | 1.1735 |
| .22026 | 0.39891 | 1.3217 | .35431 | 0.59470 | 1.3102 | .21350 | 0.77073 | 1.1672 |
| .22127 | 0.40038 | 1.3316 | .35495 | 0.59563 | 1.3377 | .21424 | 0.77181 | 1.1688 |
| .22479 | 0.40552 | 1.3460 | .35570 | 0.59673 | 1.3361 | .22567 | 0.78852 | 1.2033 |
| .22540 | 0.40642 | 1.3470 | .35611 | 0.59732 | 1.3450 | .22631 | 0.78945 | 1.2037 |
| .22607 | 0.40740 | 1.3482 | .36736 | 0.61375 | 1.3132 | .22733 | 0.79093 | 1.1885 |
| .22673 | 0.40836 | 1.3400 | .36812 | 0.61487 | 1.3026 | .23787 | 0.80633 | 1.2239 |
| .23765 | 0.42430 | 1.3696 | .36916 | 0.61639 | 1.3013 | .23848 | 0.80721 | 1.2229 |
| .23821 | 0.42513 | 1.3693 | .37000 | 0.61761 | 1.3003 | .23909 | 0.80811 | 1.2219 |
| .23887 | 0.42609 | 1.3407 | .37543 | 0.62555 | 1.2775 | .23995 | 0.80936 | 1.2206 |
| .23951 | 0.42702 | 1.3685 | .37618 | 0.62663 | 1.2798 | .24831 | 0.82158 | 1.2455 |
| .24594 | 0.43642 | 1.3429 | .37711 | 0.62800 | 1.3016 | .24888 | 0.82241 | 1.2397 |
| .24663 | 0.43742 | 1.3605 | .37785 | 0.62909 | 1.3135 | .24954 | 0.82337 | 1.2428 |
| .24728 | 0.43837 | 1.3784 | 15.12592 | 0.64281 | 1.2472 | .25020 | 0.82434 | 1.2293 |
| .24789 | 0.43926 | 1.3394 | .12712 | 0.64457 | 1.2164 | .25371 | 0.82946 | 1.2661 |
| .25456 | 0.44900 | 1.3553 | .12870 | 0.64687 | 1.2413 | .25434 | 0.83039 | 1.2652 |
| .25525 | 0.45001 | 1.3857 | .13324 | 0.65351 | 1.2040 | .25489 | 0.83118 | 1.2730 |
| .25587 | 0.45093 | 1.3872 | .13418 | 0.65488 | 1.2294 | .26543 | 0.84658 | 1.2719 |
| .25642 | 0.45172 | 1.3884 | .13501 | 0.65610 | 1.2138 | .26613 | 0.84760 | 1.2661 |
| .26341 | 0.46193 | 1.3960 | .13944 | 0.66256 | 1.2020 | .26672 | 0.84846 | 1.2772 |
| .26427 | 0.46318 | 1.3946 | .14047 | 0.66406 | 1.2116 | .26731 | 0.84932 | 1.2624 |
| .26535 | 0.46477 | 1.4124 | .14425 | 0.66959 | 1.2082 | .27022 | 0.85358 | 1.2783 |
| .26596 | 0.46565 | 1.3918 | .14491 | 0.67055 | 1.2250 | .27077 | 0.85438 | 1.2791 |
| .26965 | 0.47104 | 1.3876 | .14565 | 0.67164 | 1.2176 | .27136 | 0.85524 | 1.2886 |

TABLE 2

(Continued)

| JD HEL 2451900+ | PHASE | DMAG ($m_v$-$m_c$) | JD HEL 2451900+ | PHASE | DMAG ($m_v$-$m_c$) | JD HEL 2451900+ | PHASE | DMAG ($m_v$-$m_c$) |
|---|---|---|---|---|---|---|---|---|
| 15.27455 | 0.85990 | 1.2821 | 16.15402 | 0.14443 | 1.2528 | 16.23807 | 0.26719 | 1.1909 |
| .27543 | 0.86119 | 1.2996 | .15476 | 0.14551 | 1.2607 | .23864 | 0.26802 | 1.1918 |
| .27609 | 0.86215 | 1.2819 | .15798 | 0.15021 | 1.2684 | .24113 | 0.27166 | 1.1713 |
| .28379 | 0.87339 | 1.3032 | .15863 | 0.15116 | 1.2242 | .24186 | 0.27272 | 1.1806 |
| .28439 | 0.87427 | 1.3231 | .15932 | 0.15217 | 1.2397 | .24901 | 0.28317 | 1.2002 |
| .28533 | 0.87564 | 1.3350 | .16243 | 0.15670 | 1.2321 | .24968 | 0.28415 | 1.2014 |
| .28589 | 0.87647 | 1.3366 | .16306 | 0.15763 | 1.2505 | .25442 | 0.29106 | 1.1932 |
| .28882 | 0.88075 | 1.3471 | .16370 | 0.15856 | 1.2259 | .25949 | 0.29847 | 1.2224 |
| .28947 | 0.88169 | 1.3659 | .16647 | 0.16260 | 1.2346 | .26018 | 0.29948 | 1.2183 |
| .29010 | 0.88261 | 1.3657 | .16707 | 0.16348 | 1.2282 | .26113 | 0.30087 | 1.2129 |
| .29063 | 0.88338 | 1.3560 | .16773 | 0.16445 | 1.2306 | .26193 | 0.30203 | 1.1998 |
| .29570 | 0.89079 | 1.3891 | .17081 | 0.16894 | 1.2080 | .26520 | 0.30682 | 1.1935 |
| .29624 | 0.89158 | 1.4095 | .17144 | 0.16987 | 1.2154 | .26583 | 0.30773 | 1.2108 |
| .29690 | 0.89255 | 1.4306 | .17222 | 0.17100 | 1.1972 | .26672 | 0.30903 | 1.1943 |
| .29758 | 0.89354 | 1.4014 | .18299 | 0.18674 | 1.1614 | .26743 | 0.31006 | 1.1947 |
| .30111 | 0.89870 | 1.3810 | .18364 | 0.18769 | 1.2193 | .27055 | 0.31463 | 1.2282 |
| .30172 | 0.89958 | 1.3389 | .18428 | 0.18862 | 1.2190 | .27111 | 0.31544 | 1.2190 |
| .30232 | 0.90046 | 1.3564 | .18752 | 0.19335 | 1.2348 | .27168 | 0.31628 | 1.2098 |
| .30294 | 0.90137 | 1.3435 | .18811 | 0.19421 | 1.1927 | .27243 | 0.31737 | 1.2263 |
| .30698 | 0.90727 | 1.3535 | .18871 | 0.19509 | 1.2098 | .27565 | 0.32208 | 1.2347 |
| .30748 | 0.90800 | 1.3607 | .19295 | 0.20128 | 1.2018 | .27633 | 0.32306 | 1.2326 |
| .30808 | 0.90888 | 1.3780 | .19363 | 0.20228 | 1.1991 | .27696 | 0.32399 | 1.2393 |
| .30863 | 0.90967 | 1.4523 | .19436 | 0.20334 | 1.1961 | .28407 | 0.33437 | 1.2450 |
| 16.12292 | 0.09901 | 1.3704 | .20644 | 0.22099 | 1.1918 | .28478 | 0.33540 | 1.2293 |
| .12377 | 0.10024 | 1.3770 | .20712 | 0.22199 | 1.1813 | .28575 | 0.33682 | 1.2231 |
| .12473 | 0.10164 | 1.3733 | .20769 | 0.22282 | 1.1875 | .28651 | 0.33794 | 1.2337 |
| .12905 | 0.10795 | 1.3173 | .20841 | 0.22387 | 1.1851 | .29031 | 0.34348 | 1.2628 |
| .12988 | 0.10917 | 1.3098 | .21103 | 0.22769 | 1.1947 | .29111 | 0.34465 | 1.2396 |
| .13064 | 0.11028 | 1.3029 | .21162 | 0.22855 | 1.1884 | .29188 | 0.34578 | 1.2690 |
| .13469 | 0.11620 | 1.3153 | .21214 | 0.22931 | 1.1985 | .29267 | 0.34693 | 1.2721 |
| .13555 | 0.11745 | 1.3112 | .21272 | 0.23015 | 1.1922 | .29670 | 0.35281 | 1.2939 |
| .13645 | 0.11877 | 1.2888 | .21641 | 0.23555 | 1.1822 | .29737 | 0.35380 | 1.2828 |
| .13976 | 0.12360 | 1.2896 | .21693 | 0.23631 | 1.1878 | .29803 | 0.35476 | 1.2718 |
| .14061 | 0.12484 | 1.2897 | .21756 | 0.23724 | 1.1847 | .29871 | 0.35576 | 1.2697 |
| .14136 | 0.12594 | 1.3081 | .22570 | 0.24912 | 1.1750 | .30195 | 0.36049 | 1.2584 |
| .14461 | 0.13069 | 1.2982 | .22637 | 0.25010 | 1.1566 | .30260 | 0.36144 | 1.2563 |
| .14560 | 0.13212 | 1.2871 | .22701 | 0.25103 | 1.1627 | .30336 | 0.36255 | 1.2627 |
| .14636 | 0.13324 | 1.2677 | .22768 | 0.25201 | 1.1526 | .30412 | 0.36365 | 1.2782 |
| .14906 | 0.13718 | 1.2766 | .23256 | 0.25915 | 1.1650 | .31206 | 0.37525 | 1.2379 |
| .14966 | 0.13806 | 1.2770 | .23326 | 0.26016 | 1.1740 | .31268 | 0.37616 | 1.2486 |
| .15039 | 0.13912 | 1.2509 | .23434 | 0.26173 | 1.1834 | .31362 | 0.37753 | 1.2473 |
| .15334 | 0.14343 | 1.2713 | .23498 | 0.26268 | 1.1761 | .31422 | 0.37841 | 1.2323 |

## 3. Period Study

From my observations, the heliocentric times of minima were computed by fitting a Lorentzian function to the observed minima data points, as plotted in Figure 2 and Figure 3.
This function can be expressed as

$$y = y_0 + \frac{2A}{\pi} \frac{\omega}{4(x - x_c)^2 + \omega^2}$$

where $y_0$ is baseline offset, A is total area under the curve from baseline, $x_c$ is center of the minimum and $\omega$ if full width of the minimum at half height.

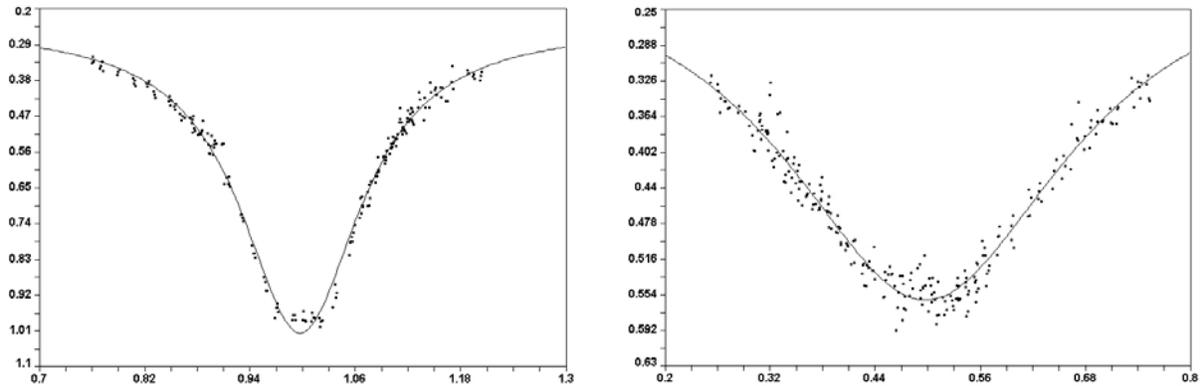

Figure 2. The primary (left) and secondary (right) minimum observations of DO Cas on filter B and the Lorentzian fits to obtain the times of minima.

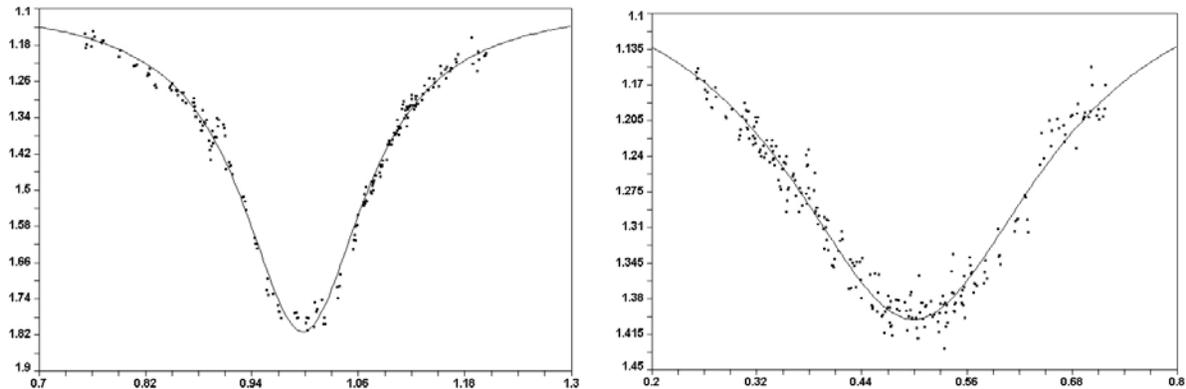

Figure 3. The primary (left) and secondary (right) minimum observations of DO Cas on filter V and the Lorentzian fits to obtain the times of minima.

The minima times, the depth of minima and the observed (O–C), in each filter, are presented in Table 4.

| | Times of minima | | Depth of minima | | O–C (day) |
|---|---|---|---|---|---|
| | Min. I | Min. II | Min. I | Min. II | |
| Filter B | 2451911.2605 | 2451910.2337 2451912.2877 | 0.64 ± 0.04 | 0.21 ± 0.06 | - 0.0020 |
| Filter V | 2451911.2606 | 2451910.2349 2451912.2889 | 0.62 ± 0.04 | 0.22 ± 0.05 | - 0.0019 |

Table 4. The new times of minima, the depth of minima and the observed (O–C), in B and V filter

The average value of the period, as is computed from both blue and yellow light of my observations, is $0^d.6846666$.

This period is nearly identical with the period of $0^d.6846661$ given by Cester *et al.* (1977) in their period study.

Seven different ephemeris formulae, including present study, have been proposed for the system DO Cas:

| | |
|---|---:|
| Min.I = HJD2428865.450 + $0^d.684655$ | Hoffmeister (1947) |
| Min.I = HJD2433282.86702 + $0^d.6846637$ | Wood and Forbes (1963) |
| Min.I = HJD2438383.6312 + $0^d.684660$ | Winkler (1966) |
| Min.I = HJD2433926.4573 + $0^d.68466595$ | Koch *et al.* (1963) |
| Min.I = HJD2439769.2130 + $0^d.68480$ | Gleim and Winkler (1969) |
| Min.I = HJD2433926.4573 + $0^d.6846661$ | Cester *et al.* (1977) |
| Min.I = HJD2451911.2606 + $0^d.6846666$ | Present Study (2001) |

It is evident from above ephemeris formulae that the orbital periods of DO Cas given by Hoffmeister (1947) and Gleim and Winkler (1969) are slightly different from those given by the other authors. Hoffmeister's epoch of primary minimum is based on the visual estimates and it may be possible that epoch is in error. The period given by Gleim and Winkler (1969) also differs at the forth decimal place, however, other periods, except the Hoffmeister's one, show consistency up to fifth decimal place.

The light curves are calculated according to the ephemeris of Koch *et al.* (1963), and the diagram of the differences between the observed and the computed (O-C) times of primary minima versus cycle numbers, presented in Figure 4, has been obtained from all times of minima published so far, according to this ephemeris. All visual, photographic and photoelectric minima were assigned a same weight in these computations.

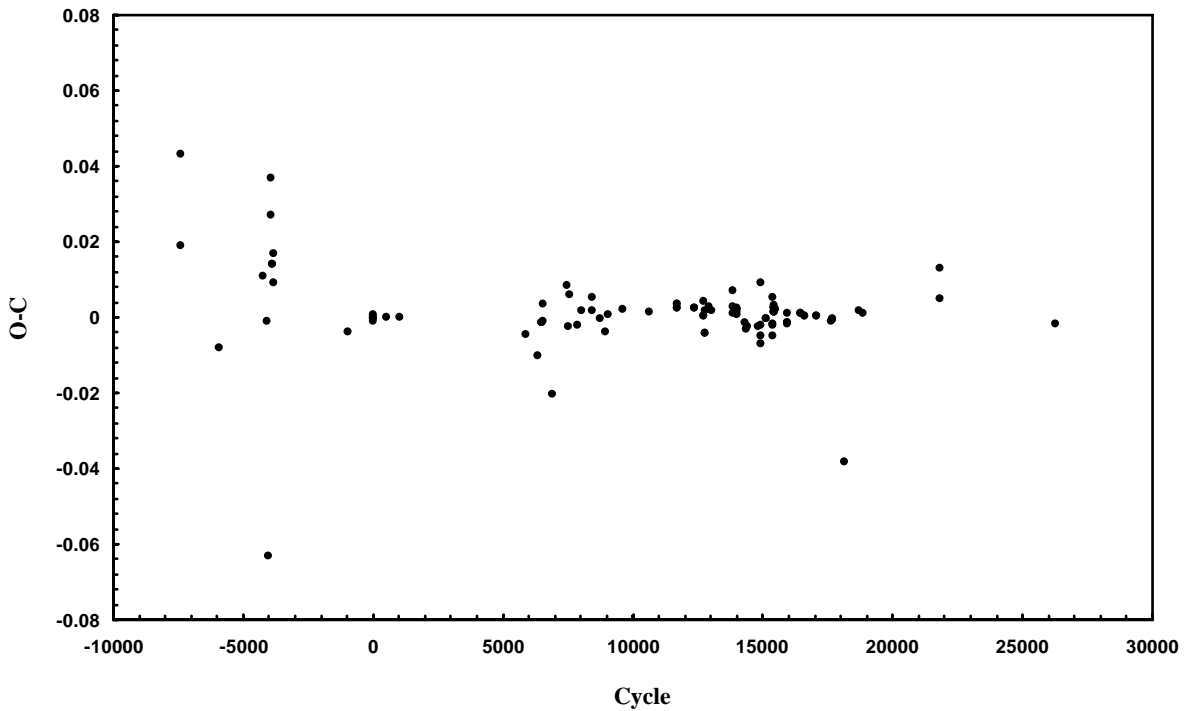

Figure 4. The O-C Diagram of DO Cas from all the times of minima

In order to study more precise the (O-C) diagram, I assigned a weight of 10 to all photoelectric times of minimum in my new computation. The current less precise measurements were weighted with a factor of 5, while the earlier visual and photographic times of minimum were totally ignored due to the large scatter in these data. Figure 5 represents the new (O-C) diagram with these times of minima which are listed in Table 5 along their O-C values.

TABLE 5

All photoelectric times of minima of DO Cas and the O-C values according to Koch *et al*. (1963) ephemeris

| JD Hel. 2400000.+ | E | O-C | Ref. | JD Hel. 2400000.+ | E | O-C | Ref. |
|---|---|---|---|---|---|---|---|
| 33926.45730 | 0 | 0.0000 | 1 | 43728.81840 | 14317 | -0.0013 | 7 |
| 33926.45752 | 0 | 0.0002 | 2 | 43777.42800 | 14388 | -0.0030 | 10 |
| 33931.93550 | 8 | 0.0009 | 3 | 43795.91440 | 14415 | -0.0026 | 7 |
| 33935.35748 | 13 | -0.0005 | 2 | 44095.79810 | 14853 | -0.0026 | 7 |
| 33937.41144 | 16 | -0.0005 | 2 | 44140.31300 | 14918 | 0.0091 | 10 |
| 34269.47501 | 501 | 0.0001 | 2 | 44142.35100 | 14921 | -0.0069 | 11 |
| 34636.45601 | 1037 | 0.0001 | 2 | 44144.41000 | 14924 | -0.0019 | 10 |
| 37960.50460 | 5892 | -0.0045 | 2 | 44146.46100 | 14927 | -0.0049 | 10 |
| 38379.52320 | 6504 | -0.0014 | 2 | 44294.35360 | 15143 | -0.0002 | 12 |
| 38383.63120 | 6510 | -0.0014 | 2 | 44451.82480 | 15373 | -0.0022 | 13 |
| 39917.28400 | 8750 | -0.0004 | 4 | 44476.47320 | 15409 | -0.0017 | 14 |
| 40051.47500 | 8946 | -0.0039 | 4 | 44477.84940 | 15411 | 0.0051 | 13 |
| 40114.46870 | 9038 | 0.0005 | 4 | 44478.52390 | 15412 | -0.0050 | 14 |
| 40518.42300 | 9628 | 0.0019 | 4 | 44485.37800 | 15422 | 0.0024 | 12 |
| 41200.34990 | 10624 | 0.0016 | 5 | 44485.37860 | 15422 | 0.0030 | 12 |
| 41936.36660 | 11699 | 0.0024 | 6 | 44498.38560 | 15441 | 0.0014 | 12 |
| 41960.33100 | 11734 | 0.0034 | 6 | 44498.38564 | 15441 | 0.0014 | 12 |
| 42405.36280 | 12384 | 0.0024 | 6 | 44498.38580 | 15441 | 0.0016 | 12 |
| 42636.77780 | 12722 | 0.0003 | 7 | 44516.87230 | 15468 | 0.0021 | 12 |
| 42636.78160 | 12722 | 0.0041 | 7 | 44830.44560 | 15926 | -0.0016 | 12 |
| 42664.84470 | 12763 | -0.0041 | 7 | 44830.44580 | 15926 | -0.0014 | 12 |
| 42776.45210 | 12926 | 0.0027 | 8 | 44859.88880 | 15969 | 0.0009 | 15 |
| 42785.35200 | 12939 | 0.0020 | 8 | 45186.47440 | 16446 | 0.0009 | 16 |
| 42787.40620 | 12942 | 0.0022 | 8 | 45306.29030 | 16621 | 0.0003 | 16 |
| 42837.38630 | 13015 | 0.0017 | 8 | 45629.45290 | 17093 | 0.0005 | 17 |
| 43408.39720 | 13849 | 0.0012 | 9 | 46001.22490 | 17636 | -0.0011 | 18 |
| 43425.51550 | 13874 | 0.0028 | 9 | 46021.08050 | 17665 | -0.0008 | 18 |
| 43425.51980 | 13874 | 0.0071 | 9 | 46021.08080 | 17665 | -0.0005 | 18 |
| 43501.51260 | 13985 | 0.0020 | 9 | 46739.29750 | 18714 | 0.0016 | 19 |
| 43502.19600 | 13986 | 0.0007 | 9 | 46831.04230 | 18848 | 0.0012 | 19 |
| 43502.19670 | 13986 | 0.0014 | 9 | 48862.45000 | 21815 | 0.0050 | 20 |
| 43512.46770 | 14001 | 0.0024 | 9 | 51911.26055 | 26268 | -0.0019 | 21 |

Ref.: 1. Koch *et al*. (1963), 2. Cester (1967), 3. Schneller and Daene (1952), 4. Pohl and Kizilirmak (1970),  5. Pohl and Kizilirmak (1972), 6. Pohl and Kizilirmak (1974), 7. Margrave (1980), 8. Cester *et al*. (1977), 9. Tumer (1978), 10. Isles (1985), 11. Rovithis-Livaniou and Rovithis (1980), 12. Pohl *et al.* (1982), 13. Margrave (1981), 14. Rovithis-Livaniou and Rovithis (1982), 15. Margrave (1982), 16. Pohl *et al.* (1983), 17. Pohl *et al.* (1985), 18. Liu *et al*. (1988), 19. Oh and Ahn (1992), 20. Aluigi *et al*. (1995), 21. Present study.

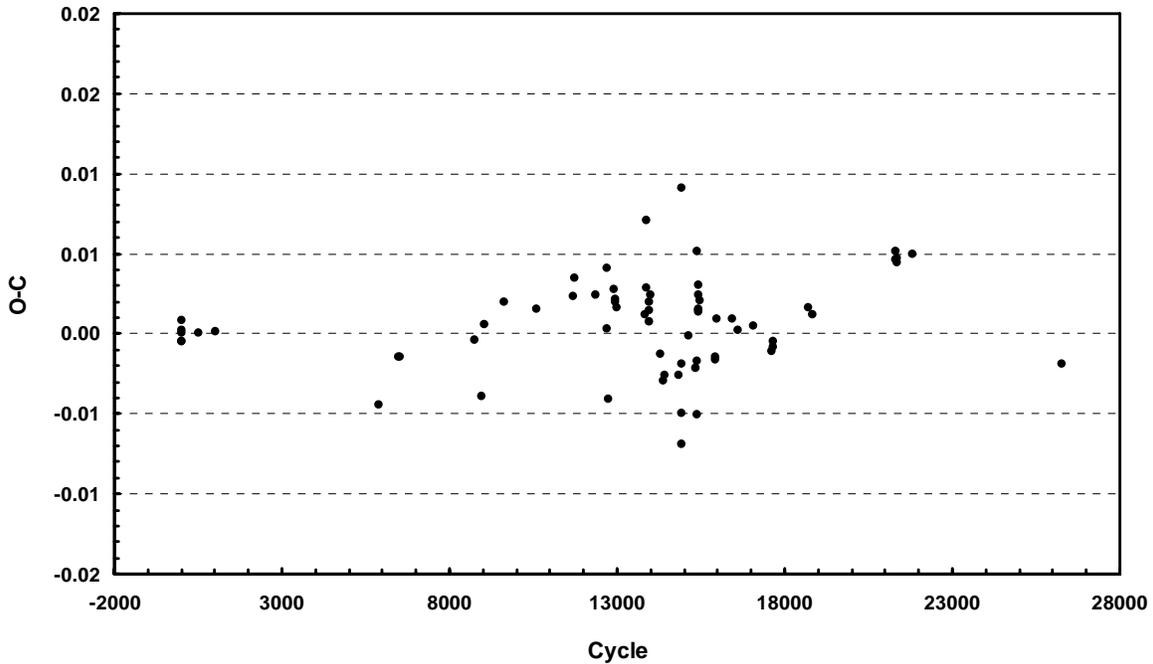

Figure 5. The O-C Diagram of DO Cas from the weighted times of minima

Since each point in the new diagram shows a small disordered scatter up to a value of $0^d.009$, the period of the system seems to be constant.

However some sine terms in ephemeris were introduced by Oh and Kim (1996), concluded that the possible explanation for these small (O-C) variation of the system DO Cas is the hypothesis of a third-body, with 0.082 solar mass and a period of 18.7 years, revolving around the close binary.

At this stage of the orbital period analysis, the possibility of the sine variation of the orbital period in DO Cas has been restudied according to the latest computed (O-C) diagram, shown in Figure 5.

Figure 6 represents this (O-C) diagram along the sine curve fitted. It's clear that the curve doesn't fit well and is totally in error.

Moreover, the square of the correlation coefficient for this fitting is 0.1924 which shows the insufficient quality for the fitting and it doesn't agree with the sine terms they suggested.

Therefore, a constant orbital period is the most possible suggestion for the system DO Cas at this moment.

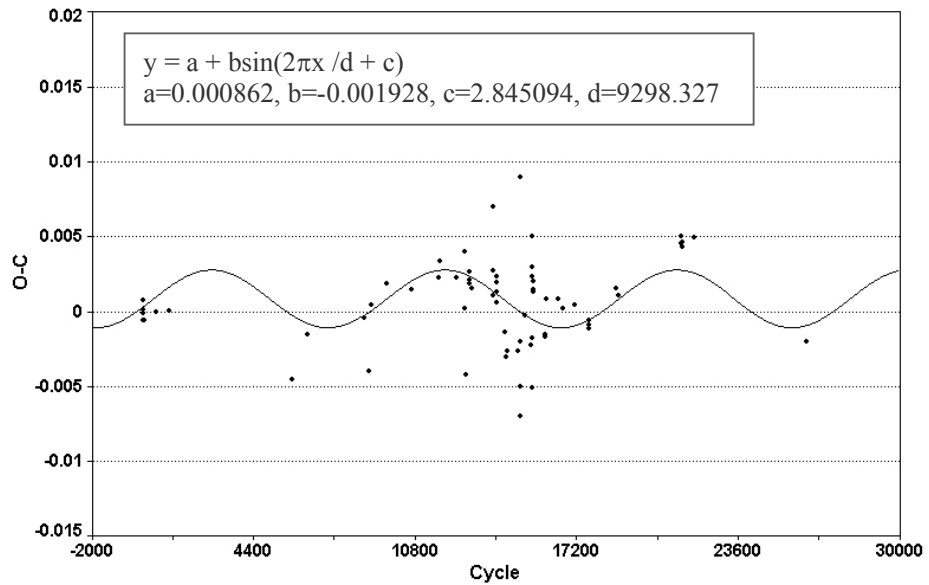

Figure 6. The O-C Diagram of DO Cas along the sine curve fitted.

## 4. Solution

The B and V light curves of DO Cas have been analyzed separately by using the latest version of Wilson code in order to derive photometric elements of this system.

The program consists of two main FORTRAN programs LC (for generating light and radial velocity curves) and DC (to perform differential corrections and parameter adjustment of LC output). The model, which the program is based on, has been described and quantified in papers by Wilson (1971, 1990, and 1998).

Since the system DO Cas had been mentioned as both a contact and a semi-detached binary by previous studies, the solution was separately performed in modes 2 (detached mode) and 3 (contact mode) of LC code. In my analysis, I assumed a value of zero for the third light as well.

DO Cas is also a single-lined spectroscopic binary (Mannino 1958), and so the spectroscopic orbit can not provide the mass ratio $q = M_2/M_1$ directly. Therefore, the solutions for several assumed values of mass ratio q (q=0.3, 0.4, 0.5, 0.6) were obtained.

Based on the spectral type A2 for the primary component of DO Cas (Henry Draper Catalogue), the temperature was estimated to be 9200K for star1 (primary). The gravity-darkening $g_1=1$ and $g_2=0.32$, and the bolometric albedo $A_1=1$ and $A_2=0.5$ were assumed according to the pervious studies. The adjustable parameters were: the inclination *i*, the mean temperature of star2, $T_2$, the monochromatic luminosity of star1, $L_1$, and the dimensionless potentials of star1 and star2, $\Omega_1$ and $\Omega_2$.

Using the LC program of Wilson, I calculated the theoretical light curves by means of the photometric elements of the best solution at q=0.315 and mode 2 and plotted them together with the observational data as shown below.

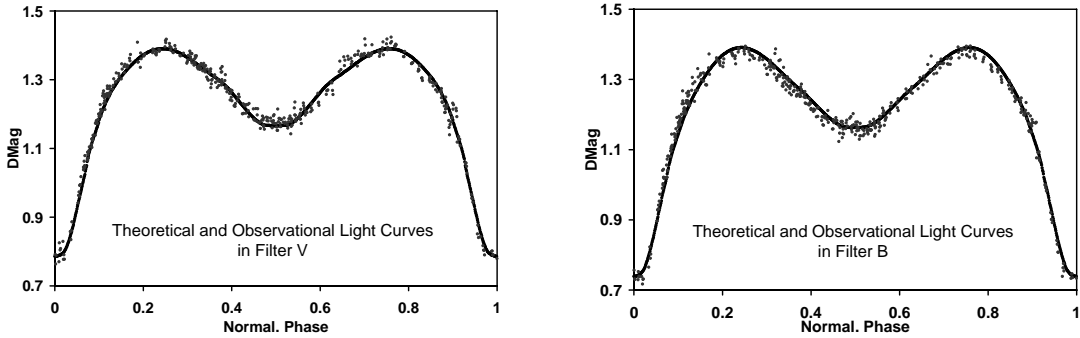

We can see that the theoretical light curves fit the observational data very well.
The final solution is given in Table 6 along the previous results by the other authors.

TABLE 6. Photoelectric elements of DO Cas from the light curves analyses

| Parameters | Present Study (2001) | Kim *et al.* (1999) | Cester *et al.* (1977) |
|---|---|---|---|
| *i* | 82.54 | 85.7 | 82.4 |
| e | 0.00 | 0 | |
| $g_1$ | 1 | 1 | |
| $g_2$ | 0.32 | 0.32 | |
| $T_1$ | 9400 K | 8200 K | 10310 K |
| $T_2$ | 5500 K | 4600 K | 5305 K |
| $\Omega_1$ | 2.5014 | 2.5008 | |
| $\Omega_2$ | 2.5014 | 2.5008 | |
| $A_1$ | 1 | 1 | |
| $A_2$ | 0.5 | 0.5 | |
| q ($m_2/m_1$) | 0.315 | 0.328 | 0.44 |
| $L_1/(L_1+L_2)_V$ | 0.947 | 0.967 | 0.979 |
| $L_1/(L_1+L_2)_B$ | 0.967 | 0.979 | 0.977 |

## 5. Discussion and Conclusions

The present study has shown that the system DO Cas is a contact binary with a large temperature difference between the two components.

To compare with the 1977 and 1999 observations published by Cester *et al.* and Kim *et al.* respectively, I also showed their results in table 6. It is very clear that the mass ratio, q, of DO Cas has changed in the time interval between 1977 an 2001 that may be explained by mass transfer which is taking place between the two components. It does agree with a B-type contact binary system which has been mentioned by some authors.

It can also be seen from the light curves in Figure 1, the Primary minimum shows a transit, whereas the secondary minimum shows an occultation.

It showed in period study that the period of the system remains constant. However Oh and Kim (1996) indicated that the orbital period of the system is changing in a sinusoidal character. There are two simple mechanisms which explain such periodic orbital period variations of eclipsing binary systems: (i) a third body, and (ii) apsidal motion. Apsidal motion is unlikely in the case of DO Cas, since the orbital eccentricity of DO Cas is obtained zero.

Therefore, the third body hypothesis could explain the sinusoidal variation of the orbital period of DO Cas, if exist. However, such a sinusoidal variation is not acceptable yet due to the insufficient fitting shown in figure 6. It might be investigated by more times of minimum to obtain a more accurate O-C diagram in future.

Unfortunately there are gaps that span many years in these data so it is difficult to follow the long term period trends. There is also some uncertainty in the exact timing of some of the eclipses; for example there were two different times of minimum for one cycle each recorded, and published which differed about an hour and twenty minutes.

Hence, at present, it is difficult to say that the third body is present in the system or not.

## Acknowledgments

I would like to thank M. R. Bagheri for his help with some of the observations. I also wish to acknowledge Professor N. Riazi, Dr. M. T. Mirtorabi as well as A. Dariush for their valuable helps in the whole process of my work.

## References

Aluigi, M., Galli G., Gaspani, A., and Venezia, P. 1995, IBVS No. 4153.
Cester, B., Giuricin, G., Mardirossian, F., and Pucillo, M. 1977, Astron. Astrophys. Suppl. 30, 223.
Dariush, A. 2001, M. Sc. Thesis, Shiraz University.
Ehersberger, J., Pohl, E., Kizilirmak, A. 1978, IBVS No. 1449.
Gleim, J. K. and Winkler, L. 1969, Astron. J. 74, 1191.
Hoffmeister, C. von. 1947, Astron. Nachr. 1, 407.
Kaluzny, J. 1985, Acta Astron. 35, 327.
Karimie, M. T. and Duerbeck, H. W. 1985, Astrophys. Space Sci. 117, 375.
Kim, H. I., Lee, W. B., Chang-Sung, E., Kyeong, J. M., and Youn, J. H. 1999, J. Astron. Space Sci. 16, 209.
Kizilirmak, A. and Pohl, E. 1971, IBVS No. 530.
Kizilirmak, A. and Pohl, E. 1974, IBVS No. 937.
Koch, R. H. 1973, Astron. J. 78, 410.
Koch, R. H., Sobieski, S., Wood, F. B. 1963, A Finding List For Observers Of Eclipsing Variables, University of Pennsylvania
Kukarkin, B. V., Kholopov, P. N., Efremov, Yu. N., Kukarkina, N. P., Kurochkin, N. E., Medvedeva, G. I., Perova, N. B., Fedorovich, V. P., and Frolov, M. S. 1969, General Catalogue of the Variable Stars,

Academy of Sciences of the U.S.S.R., Moscow.
Liu, Q. C., Zhang, Y. I., Zhang, Z. S. 1988, Acta Astron. Sinica 29, 1.
Mannino, G. 1958, Mem. Soc. Astron. Ital. 29, 433.
Margrave, T. E. 1979, IBVS No. 1631.
Margrave, T. E. 1979, IBVS No. 1694.
Margrave, T. E. 1980, IBVS No. 1869.
Margrave, T. E. 1981, IBVS No. 1930.
Margrave, T. E. 1982, IBVS No. 2086.
Margrave, T. E. Doolittle, J. H., Cutillo, D., Scherrer, J. S. 1978, IBVS No. 1478.
Oh, Kyu-Dong and Ahn, Young-Sook 1992, Astrophys. Space Sci. 187, 261.
Oh, Su-Yeon and Oh, Kyu-Dong 2000, Astron. Space Sci. 17, 163.
Oh, Kyu-Dong and Kim, Chun-Hwey 1996, Astron. Space Sci. 13, 134.
Pohl, E. and Kizilirmak, A. 1970, IBVS No. 456.
Pohl, E. and Kizilirmak, A. 1972, IBVS No. 647.
Pohl, E. and Kizilirmak, A. 1975, IBVS No. 1053.
Pohl, E., Evren, S., Tumer, O., Sezer, C. 1982, IBVS No. 2189.
Pohl, E., Hamzaoglu, E., Gudur, N., Ibanoglu, C. 1983, IBVS No. 2385.
Pohl, E., Tunca, Z., Gulmen, O., Evren, S., 1985, IBVS No. 2793.
Rovithis, P. and Rovithis-Livaniou, H. 1980, IBVS No. 1777.
Rovithis, P. and Rovithis-Livaniou, H. 1982, IBVS No. 2094.
Rovithis-Livaniou, H. and Rovithis, P. 1986, Astrophys. Space Sci. 119, 381.
Tumer, O. 1978, IBVS No. 1413.
Tumer, O. and Evren, S. 1980, IBVS No. 1863.
Wilson, R. E. and Devinney, E. J. 1971, Astrophys. J. 166, 605.
Wilson, R. E. 1990, Astrophys. J., 356, 613.
Wilson, R. E. 1998, Documentation of Computing Binary Star Observables, University of Florida
Winkler, L. 1966, Astron. J. 71, 40.
Wood, D. B. and Forbes, J. E. 1963, Astron. J. 68, 257.